\documentclass[reprint,secnumarabic,amssymb,amsmath,aps,prb,groupedaddress,frontmatterverbose,showpacs]{revtex4-1}
\usepackage[dvips]{graphicx}
\usepackage{bm}
\usepackage{color}
\begin{document}
\title{Magnetic states of Ni-Mn-Sn based shape memory alloy: a combined muon spin relaxation and neutron diffraction study}
\author{J. Sannigrahi$^1$, S. Pramanick$^2$, S. Chatterjee$^3$, J. S. Lord$^1$, D. Khalyavin$^1$, A.D. Hillier$^1$, D. T. Adroja$^{1,4}$,  S. Majumdar$^3$}
\email{sspsm2@iacs.res.in}
\affiliation{$^1$ISIS Neutron and Muon Source, Science and Technology Facilities Council, Rutherford Appleton Laboratory, Didcot
OX11 0QX ,United Kingdom}
\affiliation{$^2$UGC-DAE Consortium for Scientific Research, Kolkata Centre, Sector III, LB-8, Salt Lake, Kolkata 700 098, India}
\affiliation{$^3$School of Physical Sciences, Indian Association for the Cultivation of Science, 2A \& B Raja S. C. Mullick Road, Jadavpur, Kolkata 700 032, India}
\affiliation{$^4$Highly Correlated Matter Research Group, Physics Department, University of Johannesburg, Auckland Park 2006, South Africa}

\begin{abstract}
The fascinating multiple  magnetic states observed in the  Ni-Mn-Sn based metamagnetic shape memory alloy are addressed  through a combined muon spin relaxation ($\mu$SR) and neutron powder diffraction studies. The material used in the present investigation is an off-stoichiometric alloy of nominal composition, Ni$_{2.04}$Mn$_{1.4}$Sn$_{0.56}$. This prototypical alloy, similar to other members in the Ni-Mn-Sn series, orders ferromagnetically below $T_{CA}$ (= 320 K), and undergoes martensitic type structural transition at $T_{MS}$ (= 290 K), which is associated with the sudden loss of magnetization. The sample  regains its magnetization below another magnetic transition at $T_{CM}$ = 260 K. Eventually, the composition shows a step-like anomaly at $T_B$ = 120 K, which is found to coincide with the blocking temperature of exchange bias effect observed in the alloy.  In our study, the initial asymmetry ($A_{10}$ ) of the $\mu$SR data falls rapidly below  $T_{CA}$, indicating  the onset of bulk magnetic order. $A_{10}$ regains its full asymmetry value below  $T_{MS}$  suggesting the collapse of the ferromagnetic order into a fully disordered paramagnetic state. Below the second magnetic transition at $T_{CM}$, asymmetry drops again, confirming the re-entrance of a long range ordered state. Interestingly,  $A_{10}$ increases sluggishly below $T_B$, indicating that the system attains a disordered/glassy magnetic phase below $T_B$, which is responsible for the exchange bias and frequency dispersion in the ac susceptibility data as previously reported. The neutron powder diffraction data  do not show any magnetic superlattice reflections, ruling out the possibility of a long range antiferromagnetic state at low temperatures. The ground state is likely to be comprised of a concentrated metallic spin-glass  in the backdrop of an ordered ferromagnetic state.

\end{abstract}
\maketitle

\section{Introduction}
Advanced functional materials play an ever increasing role in the modern technological developments, which encompass areas such as energy harvesting, computation, communication as well as to combat environmental pollution. Broadly, they can be classified into five groups depending upon their functionality; Adaptive, Magnetic, Electric, Optical, and Energy and Environmental materials. It is needless to mention that a proper investigation of their physical properties is important in understandings the essential physics associated with their functionality. For example, the study of CuO-based high $T_C$ superconductors has enriched our understanding of electronic properties of correlated  metal oxides. It is to be noted that many of these functional materials are actually multifunctional, {\it i.e.}, they show two or more functional properties. Ni$_{2.04}$Mn$_{1.4}$Sn$_{0.56}$, the title composition of this work, is one such material having functionality as adaptive, magnetic, electric as well as energy and environmental material. 

\par
Ni$_{2.04}$Mn$_{1.4}$Sn$_{0.56}$ belongs to a class of materials known as  metamagnetic shape memory alloys (MSMAs). They show large magnetic field induced strain, magneto-resistance, magneto-caloric, baro-caloric, and exchange bias effects.~\cite{kainuma1, krenke, kainuma2, manosa, chatterjee,barocaloric} The general compositions of these alloys can be expressed as Ni$_2$Mn$_{1+p}$Z$_{1-p}$ (Z= In, Sn, Sb and $p <$ 1), and the observed functionality arises from their bi-ferroic nature with the simultaneous presence (as well as their mutual interplay) of ferromagnetism and ferroelasticity. The alloys are characterized by first-order martensitic type structural phase transition occurring at a temperature $T_{MS}$ and a second order paramagnetic (PM) to ferromagnetic (FM) transition at the Curie point ($T_C$). For the practical realization of magnetic shape memory and other magneto-functional properties, one should have  $T_C > T_{MS}$, {\it i.e.}, the sample should be in a magnetically ordered state when martensitic phase transition (MPT) takes place.        

\begin{figure}[t]
\begin{center}
\includegraphics[width = 8 cm]{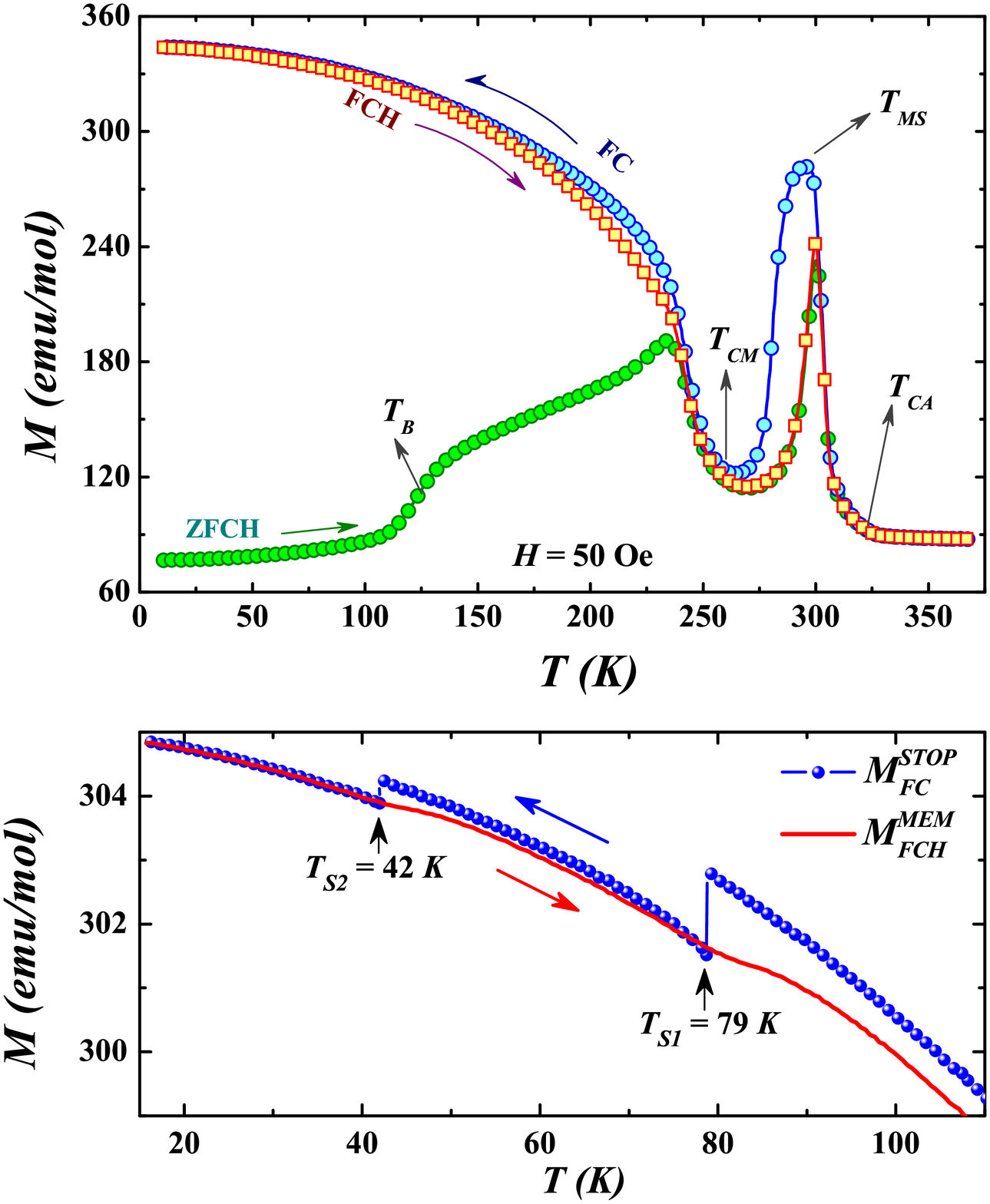}
\caption {In the upper panel, magnetization is plotted as a function of temperature in zero-field-cooled heating (ZFCH), field cooling (FC) and field-cooled heating (FCH) conditions in presence of 50 Oe of external magnetic field . The martensitic transformation ($T_{MS}$) is clearly visible around 300 K, where large thermal hysteresis is present. $T_{CA}$, $T_{CM}$ and $T_B$ indicate austenite Curie point, martensite Curie point and exchange bias blocking temperature respectively. The lower panel shows the field-cooled-field-stop memory data indicating the presence of glassy magnetic phase below 120 K. Here $M_{FC}^{STOP}$ denotes the magnetization data recorded during cooling along with 1 h stops at 80 K and 40 K. Magnetic field was turned off during the stops. $M_{FCH}^{MEM}$ denotes the subsequent heating run, and clear anomalies (in the form of shallow dips) were observed at the stopping points signifying the presence of memory.}
\end{center}
\end{figure}
     
\par
The presence of two critical temperatures (namely $T_C$ and $T_{MS}$) makes the system to have a rather exotic phase diagram. There are few important aspects associated with these alloys, which remain elusive till date. Firstly, what happens to the FM  state of the alloy (we are considering $T_C > T_{MS}$) right below the MPT? The high temperature structural phase is called austenite with cubic lattice symmetry, while the low temperature phase (below $T_{MS}$) is called martensite with a tetragonal/orthorhombic/modulated structure. It is found that magnetization drops sharply below  $T_{MS}$, indicating the loss of ordered FM moment. Some diffuse peaks were  observed in the Neutron Powder Diffraction (NPD) data of Ni-Mn-Sn alloys, which were assigned to the existence of incipient antiferromagnetic (AFM) coupling~\cite{brown}. As evident from the chemical compositions, a fraction of Z atoms is replaced by Mn (we call it Mn$^{\prime}$), and theoretical calculations indicate the  existence of AFM  correlation between regular Mn and Mn$^{\prime}$ atoms.~\cite{montecarlo,xps,tang,vvs-th,hbx-th,cmli-th,pirolkar} On the other hand, $^{57}$Co-Rh M\"{o}ssbauser spectroscopy on a Ni-Mn-Sn alloy (with small amount of enriched $^{57}$Fe doped at the Mn site) indicated a PM state below $T_{MS}$ of the sample.~\cite{umetsu} NPD studies, performed on various Ni-Mn-Z alloys, fail to identify well defined magnetic Bragg peaks associated with an ordered AFM state.~\cite{brown,brown-In} Neutron polarization study indicates the existence of FM correlations, which vanishes below  $T_{MS}$ with the concomitant occurrence of Mn-Mn$^{\prime}$ AFM correlations.~\cite{aksoy} Therefore, the nature of the magnetic state just below $T_{MS}$ remains uncertain; it may be  incipient AFM in the backdrop of a PM/FM phase, a long range ordered AFM state with weak moment or an FM state with reduced Mn moment.

\par
The second unresolved point is associated with the magnetic ground state of these alloys. The off-stoichiometric Ni-Mn-Z alloys show a step-like feature well below $T_{MS}$ in the zero-field-cooled magnetization data and exchange bias (EB) effect was observed below $T_B$. Our group previously reported that field-cooling from just above $T_B$ is sufficient to observe EB, and  $T_B$ actually signifies a spin freezing temperature.~\cite{souvik2} Subsequently, there have been numerous reports on the glassy magnetic state of the Ni-Mn-Z alloys.~\cite{cong,wang,umetsu1,sangam} Nevertheless, ambiguities remain on the nature of the glassy state, and the ground state has been described as re-entrant spin glass, cluster glass or super spin glass by various authors (cited in the previous line). EB generally requires two different magnetic phases (say, FM $+$ AFM or spin-glass $+$ FM) to be present. Presence of AFM clusters below $T_{MS}$  is highly possible, since there is a strong evidence for incipient antiferromagnetism. It remains unclear whether the ground state is characterized by a (i) mixture of AFM and FM phase fractions along with interfacial glassyness, (ii) coexisting spin-glass (SG) and FM phases, or (iii) stand alone SG phase.

\par
In the present work, we have addressed  these  aspects using muon spin resonance/rotation ($\mu$SR) as well as  NPD techniques on an MSMA of  nominal composition, Ni$_{2.04}$Mn$_{1.4}$Sn$_{0.56}$. One can notice that a small amount of excess Ni is doped at the expense of Sn, which is required to elevate $T_{MS}$. While ferromagnetic Curie point of the high temperature austenite is around $T_{CA}$ = 320 K, the structural transition takes place around $T_{MS}$ = 290 K. The reason for choosing this composition lies with the fact  that $T_{MS}$ is very close to $T_{CA}$, and the FM state becomes unstable below the MPT (see fig. 1). The sample  regains its magnetization below a second transition at $T_{CM}$ = 260 K  in the martensitic phase. The step like anomaly is seen below $T_B$ = 120 K, and considerable EB is observed at low temperature.~\cite{sabya} The sample shows the signature of field-cooled-field-stop memory (FCFS)~\cite{sabya1} as shown in the lower panel of fig. 1. The observed FCFS in this bulk sample indicates the presence of glassy magnetic ground state. 

\begin{figure}[t]
\begin{center}
\includegraphics[width = 8 cm]{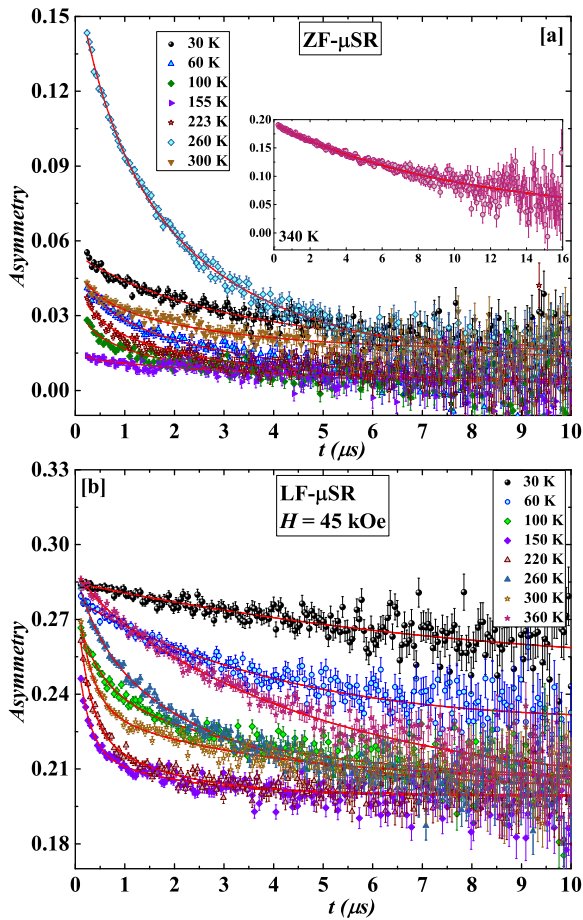}
\caption {(a) shows the muon spin relaxation data measured in zero magnetic field at different constant temperatures. The inset shows the same recorded at 340 K, where the sample is in the paramagnetic austensitic phase. (b) depicts the similar data recorded in presence of 45 kOe of applied magnetic field.}  
\end{center}
\end{figure}    
      
\section{Experimental Details}  
The polycrystalline sample of Ni$_{2.04}$Mn$_{1.4}$Sn$_{0.56}$  for the present study was  prepared by argon arc melting the constituent elements.~\cite{sabya}  The temperature ($T$) dependent dc magnetization ($M$) measurements were performed  using a commercial Quantum Design SQUID magnetometer (MPMS 3). The $\mu$SR measurements were performed at ISIS facility, Rutherford Appleton Laboratory, UK using EMU (for zero magnetic field) and HIFI (for longitudinal magnetic field) spectrometers. The sample was mounted on a silver sample holder to minimize the background and measurements were performed at different temperatures. The neutron powder diffraction was carried out at the WISH time of flight diffractometer at the ISIS facility between 8 and 363 K.  The powdered sample was inserted in a cylindrical vanadium container of 6 mm diameter. A standard Helium closed cycle refrigerator was used to cool the sample down to 8 K. Nuclear and magnetic structure refinements were performed by the Rietveld method using the FULLPROF program.~\cite{fp} The diffraction peaks were described by a pseudo-Voigt profile function. 

\section{Results}

\subsection{Muon Spin Relaxation}
$\mu$SR is an accomplished local probe technique to study the magnetism of a material.~\cite{blundell} Spin polarized positive muons ($\mu^{+}$) are implanted into the sample. The implanted muons decay into positrons, which are emitted preferentially in the direction of the muon spin. The spin of the implanted $\mu^{+}$ precesses around the effective magnetic field vector at the site of implantation. Random and fluctuating fields within the sample can depolarize the muons. In the actual zero field (ZF) or longitudinal field (LF) experiment, the emitted positrons are counted parallel and anti-parallel to the initial muon spin direction. The difference between the number of positrons in the forward and backward directions is generally measured as a function of time ($t$), and it is called the asymmetry function, $G_z(t)$. Since, $G_z(t)$ is a measure of muon depolarization, one can get significant information on the magnetic state of the material out of it.~\cite{ryan,dalmas}

\begin{figure}[t]
\begin{center}
\includegraphics[width = 9 cm]{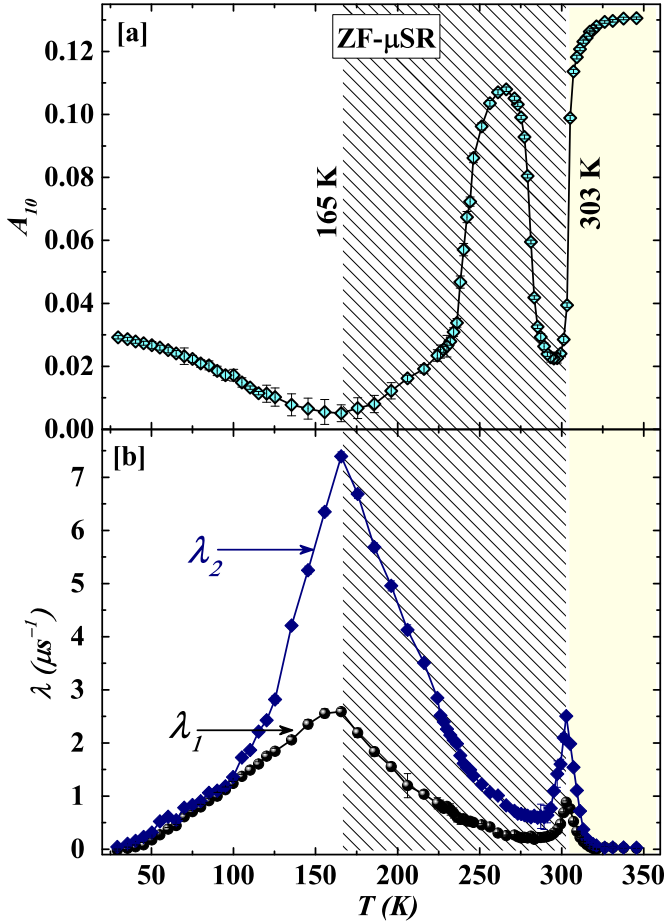}
\caption {(a) shows the muon asymmetry, $A_{10}$ as a function of temperature measured in zero magnetic field while cooling. (b) shows the thermal variation of muon depolarization rate, $\lambda_1$ and $\lambda_2$ at $H$ = 0.}
\end{center}
\end{figure}
\par
For a typical magnetic material, the relaxing part of the asymmetry often obeys an exponential law, $G_{z} = A_0\exp{(-\lambda t)}$, where $A_0$ is the initial asymmetry, and $\lambda$ is the (spin-lattice) relaxation rate. For the present Ni-Mn-Sn alloy, the simple exponential law is found to be inadequate to describe the data at all temperatures. The simplest approach is to have a  bimodial distribution with two relaxation rates, $\lambda_1$ and $\lambda_2$, which results,~\cite{manganite1}  
\begin{equation}
\centering
G_{z}(t) = A_{10}\exp{(-\lambda_1 t)}  +  A_{20}\exp{(-\lambda_2 t)} + b_g
\label{bimodial}
\end{equation}
Here $b_g$ denotes the time independent background of asymmetry. Similar two exponent model has been used for diverse magnetic systems successfully, which include perovskite manganites,~\cite{manganite1,kawasaki,heffner} cobaltates,~\cite {guo}, as well as Hesuler based intermetallic alloys.~\cite{hiroi}  Muons, whose nearest atom on the Z sublattice is either Mn$^{\prime}$ or Sn, are subject to  different internal fields. The regression converges better over full $T$ range, if we put the constrain $A_{20} = 0.4A_{10}$, and the fittings  presented in the subsequent parts have been performed considering the above constrain. The value of $b_g$ for a particular fixed magnetic field  was estimated from $T$ = 30 K data and it  is kept constant at all $T$ for a particular value of  $H$. For the ZF case, the value of $b_g$ was kept constant at 0.0094(4), while for the LF data it is fixed at the value 0.1973(4).

\par
Considering glassy ground state in the studied alloy, we have additionally used a stretched exponential function to fit the data below 150 K.~\cite{ryan,stretch1,nav2o5,mncosi,pbfenbo}

\begin{equation}
\centering
G_{z}(t) = A_{10}\exp{[(-\lambda t)^{\beta}]}  + b_g
\label{stretch}
\end{equation}

Here $\beta$ is called the shape parameter and $b_g$ is the background. Similar to the double exponential fitting, we have kept $b_g$ fixed for all $T$ values for a particular $H$. 
 
\par
Fig. 2 (a) shows the time domain $\mu$SR data recorded at different temperatures in ZF condition while cooling.  At $T$ = 340 K, the asymmetry shows an exponential like decay, which is expected for a PM state. On lowering the temperature below the FM Curie point ($T_{CA}$ = 320 K), $G_{z}(t)$ shows a fast relaxation component, which coexists with the slow relaxing part. This is clearly due to the onset of FM transition in the system. On further lowering $T$ below $T_{MS}$ (= 290 K), the fast relaxation component appears to get diminished.  This corresponds to the rapid fall of $M$ below $T_{MS}$ in the magnetization data [see fig. 1 (a)]. The fast relaxation reappears when the sample is cooled below the second magnetic ordering point $T_{CM}$. Most interestingly, the damping gradually  gets reduced, when the sample is cooled below the blocking temperature, $T_B$ = 120 K. We also recorded the $\mu$SR spectra in presence of 45 kOe of applied longitudinal field, as depicted in fig. 2(b).

\begin{figure}[t]
\begin{center}
\includegraphics[width = 9 cm]{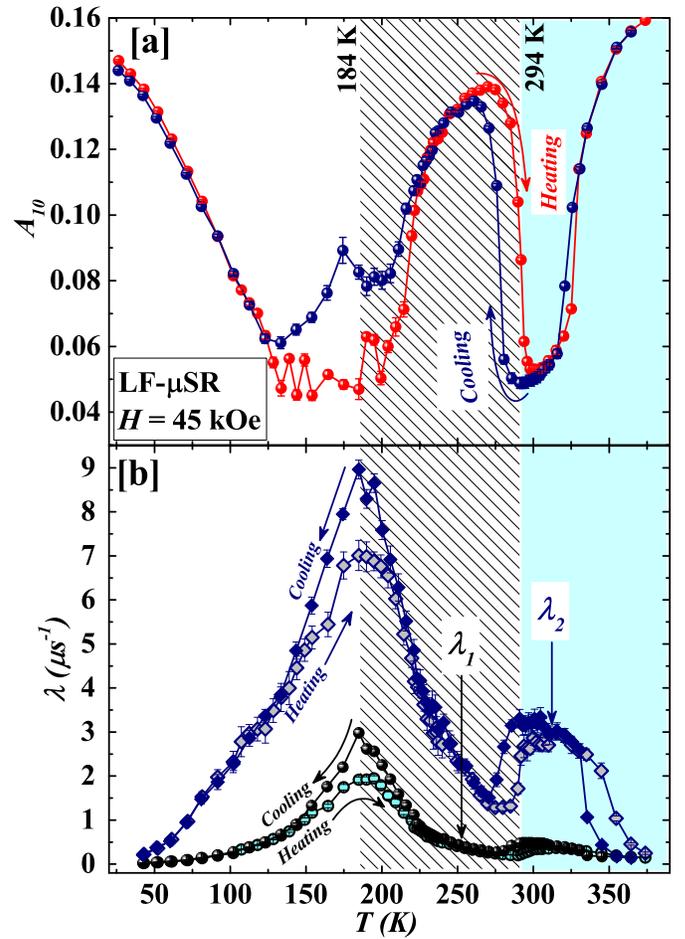}
\caption {(a) shows the muon asymmetry, $A_1$ as a function of temperature measured in 45 kOe of longitudinal magnetic field. (b) shows the thermal variation of muon depolarization rate, $\lambda_1$ and $\lambda_2$ at $H$ = 45 kOe.}
\end{center}
\end{figure}

\par
In order to  elucidate the magnetic state of the sample at different $T$'s, we have fitted the time domain data with the double exponential function stated in eqn.~\ref{bimodial}. The solid lines in figs. 2(a) and (b) represent the fit to the data. The values of $A_{10}$ and  ($\lambda_1$, $\lambda_2$), obtained by  fitting the ZF data, are plotted in figs. 3(a) and (b) respectively. 

\par
The $T$ variation of $A_{10}$ [see fig.3 (a)] provides noteworthy information on the magnetic state of the studied alloy. In the PM state (above 310 K), the initial asymmetry is found to be almost $T$ independent.  On cooling, $A_{10}$ falls sharply with the onset of ferromagnetism at $T_{CA}$, and it attains its lowest value just below 300 K. The structural transition is beautifully echoed in that data, as $A_{10}$ again rises sharply below $T_{MS}$ and attains a value ($=$ 0.11) slightly lower than the value in the high-$T$ PM state ($=$ 0.13). Below 260 K, $A_{10}$ shows another sharp fall, which can easily be assigned to the second magnetic transition occurring at $T_{CM}$. The most remarkable observation is the sluggish rise (as opposed to the sharp change at magnetic Curie points) of $A_{10}$ with decreasing $T$ below  $T_B$. This is an indication for the loss of magnetic order, and it nicely fits with the conjecture of a glassy magnetic ground state of the system. It is  evident that even at the lowest temperature, the value of $A_{10}$ ($=$ 0.03) is much smaller than the value observed in the high-$T$ PM state. The $T$ variation of $\lambda_1$ and $\lambda_2$ are plotted in fig. 3 (b). Both the parameters show well defined peaks at the long range magnetic ordering temperatures, which corroborates well with the nature of the $A_{10}(T)$ curve. It is  worthwhile to mention that across the  transition from PM to a magnetically ordered phase, one would expect the relaxing part of asymmetry to decrease by $\frac{1}{3}$ for a powder/polycrystalline sample in ZF measurement. The residual $\frac{2}{3}$ component in the time domain data will correspond to muon's damped oscillations due to the internal magnetic field.~\cite{ryan,dalmas} In the present case, the frequency of oscillation may be too high too be observed in the spectrometer at ISIS due to the large internal field from Mn.

\par
We have  fitted the LF relaxation curves recorded under 45 kOe, and the $T$ variation of resulting parameter $A_{10}$ is depicted in fig. 4(a). For the LF data, we recorded relaxation during both heating and cooling. The variation of LF $A_{10}(T)$ traces similar nature as that of ZF one. However, it provides few pieces of additional information, which are not obvious in the ZF data. Firstly, the sharp rise in $A_{10}$ due to MPT gets shifted to a lower temperature under LF. It is well known that the magnetic field favors austenite and it reduces $T_{MS}$, which gets well reflected in our $\mu$SR data. Secondly, clear thermal hysteresis is seen in $A_{10}(T)$ around the martensitic transition occurring at $T_{MS}$. The hysteresis is  present in the plot of $\lambda_1$ and $\lambda_2$  as well [fig. 4 (b)], and it can be  accounted by the first order nature of the structural transition. A second thermal hysteresis is  present just below $T_{CM}$, which can be traced back to the similar thermal hysteresis observed in the bulk magnetization data (see fig. 1). Such hysteresis may be linked with the first order nature of the magnetic transition at $T_{CM}$. 

\begin{figure}[t]
\begin{center}
\includegraphics[width = 8.5 cm]{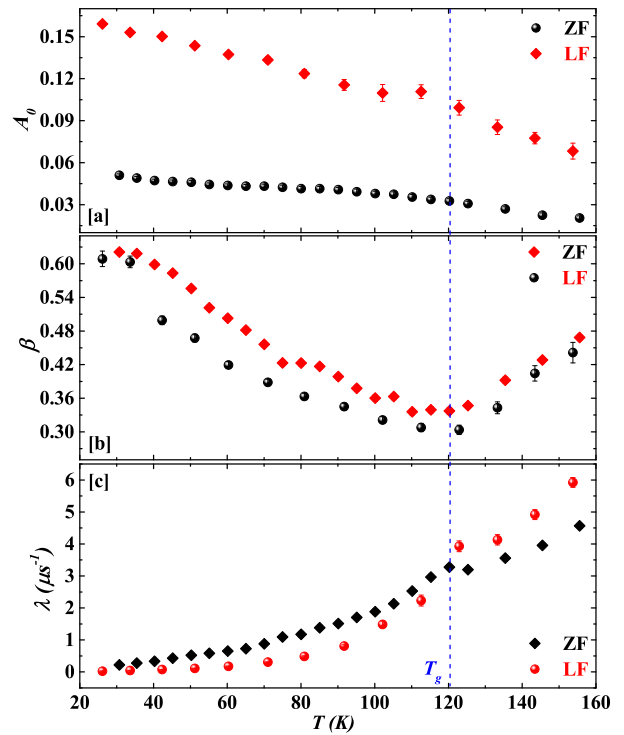}
\caption {(a), (b) and (c) respectively represent the temperature variation of  initial asymmetry ($A_0$), exponent ($\beta$) and relaxation rate ($\lambda$) as obtained from fitting the time domain data using stretched exponential function as described in eqn.~\ref{stretch} Both ZF and LF (50 kOe)  data are represented in the plots.}
\end{center}
\end{figure}

\par
Apart from the double exponential fitting (eqn.~\ref{bimodial}), we have  used the stretched exponential function (as described by eqn.~\ref{stretch}) to fit the time domain data. A stretched exponential form of muon depolarization is generally expected both above and below the spin freezing temperature ($T_g$). For a simple PM to SG transition, $\beta$ attains a value close to unity at a temperature well above $T_g$. On cooling towards $T_g$, $\beta$ decreases.~\cite{mncosi} It has been found that for so-called concentrated canonical spin glasses (where there is a distribution of the frequency of fluctuation of the local magnetic field), $\beta$ attains a value of $\frac{1}{3}$ at $T_g$.~\cite{stretch1}     

\par
We find that both the ZF and LF relaxation data can also be  fitted well  with a stretched exponential function as described in eqn.~\ref{stretch} below about 150 K. In fact the quality of the fittings in the temperature range 30-150 K  is found to be better (as evident from the lower values of $\chi^2$ of the fits) in case of stretched exponential  as compared to two exponential function. However, stretched exponential fitting  turns poorer above 150 K, and it fails to converge with physically meaningful values of the fitting parameters. In figs. 5 (a), (b), and (c), we have shown the $T$ variations of the parameters $A_0$, $\beta$and $\lambda$ respectively  only below 150, which were obtained by fitting the ZF and LF relaxation data recorded while the sample is being  cooled. The initial asymmetry ($A_0$) shows a rise below 150 K, similar to  the behavior of $A_{10}$ obtained from the two exponential model. The exponent $\beta$ decreases as we approach $T_B$ from high temperature, and shows a minimum at 120 K [fig. 3 (c)]. The values $\beta$ at $T_B$ are found to be $\beta$(ZF)$_{T_B}$ = 0.33(4) and $\beta$(LF)$_{T_B}$ = 0.30(6). These values are fairly close to the $\beta$ = $\frac{1}{3}$  law (particularly the ZF one) for concentrated metallic SG. Below $T_B$, it rises again and attains a value of 0.62 at 30 K.  On the other hand, $\lambda$ shows a decreasing trend on cooling along with peak like feature at $T_B$.

\begin{figure*}[t]
\begin{center}
\includegraphics[width = 16 cm]{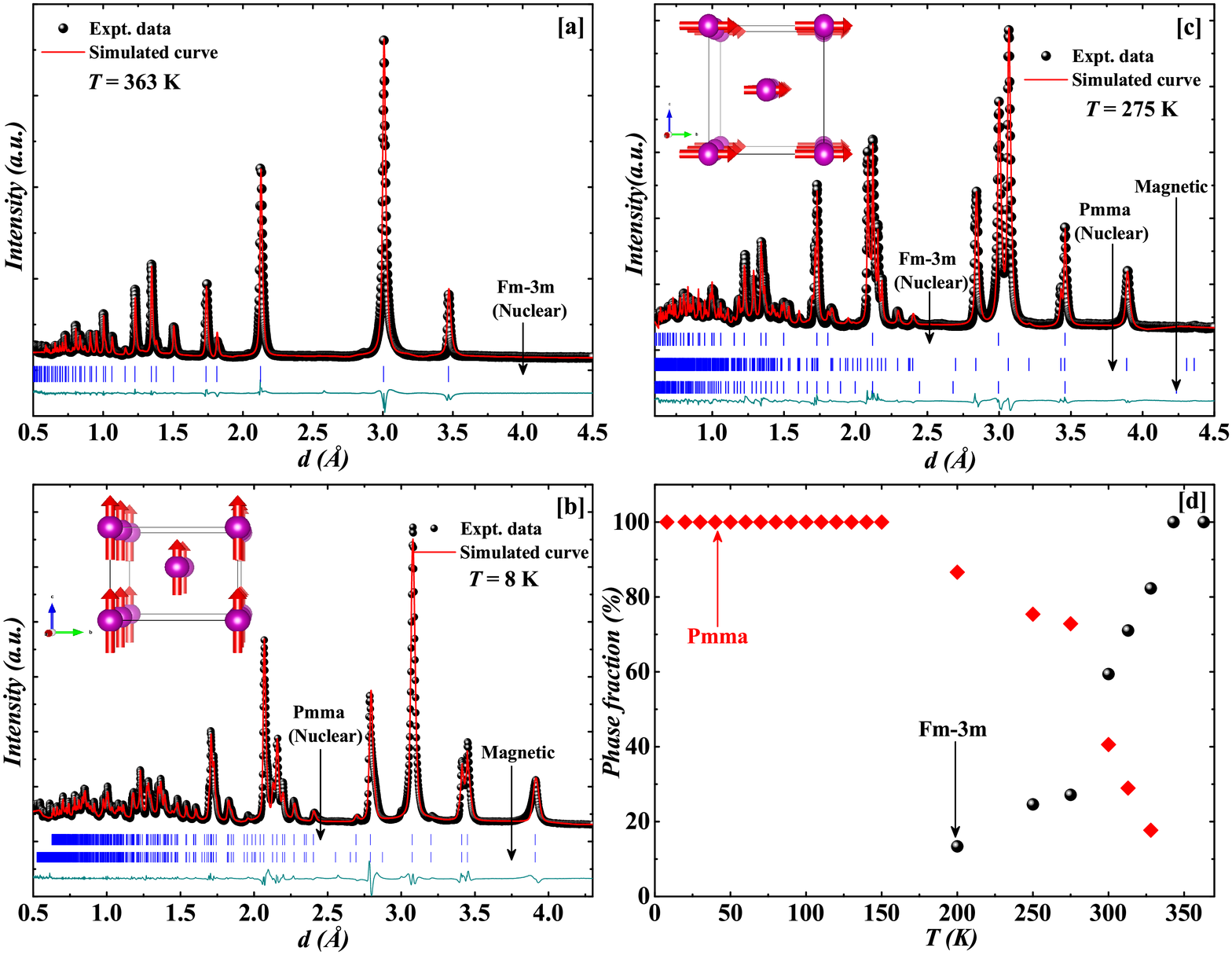}
\caption {(a), (b) and (c) show the powder neutron diffraction data recorded at 363 K, 8 K and 275 K respectively. The solid lines are  simulated  by using the FullProf suite for Rietveld refinements. (d) shows the temperature variation of martensite and austenite phase fractions as obtained from the  Rietveld refinement of the neutron diffraction data.}
\end{center}
\end{figure*}

\subsection{Neutron powder diffraction measurements}
Figs. 6 (a), (b) and (c) show the high resolution NPD data measured at different temperatures. The sample was first heated to 363 K (which is well above $T_{MS}$ and $T_{CA}$) and diffraction data were recorded while cooling from 363 to 8 K within the closed cycle refrigerator. The diffraction pattern at 363 K can be well indexed by the cubic L2$_1$ structure with space group $Fm\bar{3}m$ as  expected for a pure austenitic phase [see fig. 6 (a)]. At 363 K, the sample is in the PM state, and a good refinement is obtained by considering only the nuclear contribution of the cubic austenite phase with L2$_1$ geometry. The refined cubic lattice parameter is found to be $a_c$ = 5.991(1) \AA. 
On cooling below $T_{MS}$, the cubic peaks start to disappear, along with the appearance of martensitic peaks [see the NPD data at 275 K in fig. 5 (c)]. 

\begin{table}
Magnetic moment fixed along [010] direction \\
			$\chi^2$ of fit = 6.15\\
			\begin{tabular}{ p{1cm} p{1cm} p{1cm} p{1cm} p{1cm}  p{1cm}}
					\hline \hline
 Atom &  Site & X & Y & Z & moment ($\mu_B$) \\
							\hline 
							
 Mn & 4$a$ & 0 & 0 & 0 & 1.780(7) \\ \\
\hline
 Mn & 4$b$ & 0.5 & 0.5 & 0.5 & 1.131(5) \\ \\

\hline		
		\end{tabular}
		
	\caption{Magnetic structure data for the cubic austenite ($Fm\bar{3}m$) at 275 K as obtained from the powder neutron diffraction measurements.}
	\label{mag275}
\end{table}

\par
     
\par
At 8 K, the data can be described by a single orthorhombic phase as shown in fig. 6 (b). We do not observe any well resolved magnetic superlattice reflection, which matches well with the previous report.~\cite{brown} This rules out the possibility of an ordered AFM state below $T_{MS}$. In case of related Ni-Co-Mn-Ga based Hesuler alloys, distinct AFM state was observed below the MPT.~\cite{orlandi}  A stable refinement is achieved assuming the nuclear phase coming from $Pmma$, and one magnetic phase with propagation vector $k$ = (0, 0, 0). Here we have assumed that the ordered moments arise from the Mn atoms residing at 2$a$ and 2$f$ positions only, and neglected any contribution from Ni. We have fitted our data with several possible options of collinear magnetic structure with $k$ = (0, 0, 0), and the best fit is obtained when the moments are aligned along orthorhombic $c$~axis.  Mn atoms, situated at 2$a$ and 2$f$ sites, have ordered moment 2.76 $\mu_B$ and 2.30 $\mu_B$ respectively (see Table~\ref{mag8}).

\par   
In fig 6. (c), we have plotted NPD data for an intermediate  temperature of 275 K, where both  $Fm\bar{3}m$ (austenite) and $Pmma$ (martensite) phases coexist, and we have considered both the phases to refine the diffraction data. As evident from our $\mu$SR data, the asymmetry rises below $T_{MS}$, indicating a PM martensitic phase. However, the residual austenite fraction may still be present in the sample having an ordered FM state. Our effort to fit 275 K data  considering only the nuclear contributions coming from cubic and orthorhombic phases do not provide a good convergence.  A better fit is obtained, when the  FM contribution from the cubic phase is taken into consideration. Fig 6 (c) shows the experimental data as well as the refinements. The ratio of the volume fraction of the cubic and orthorhombic phases is found to be $\frac{2}{3}:\frac{1}{3}$. This indicates that below $T_{MS}$, the major phase fraction is martensite, although a sizable austenite phase is still present. The cubic and orthorhombic lattice parameters are found to be $a_c$ = 5.991(7) \AA~and $a_o$ = 8.613(4) \AA~,   $b_o$ = 5.675(6) \AA~, $c_o$ = 4.360(5) \AA~respectively. The ordered Mn moments are found to be 1.78 and 1.13 $\mu_B$ at 4$a$ and 4$b$ sites respectively (see Table~\ref{mag275}). These moment values match quite well with the previous report.~\cite{brown}
  
\par
Interestingly, a fraction of high-$T$ austenite  continues to exist over a wide temperature range well below $T_{MS}$. Eventually, the reflections due to austenite  disappears  when the sample is cooled below 200 K. In order to determine the $T$ variation of  phase fraction, we have performed structural refinements of the NPD data  at different temperatures between 8 K and 363 K. Considering the coexistence of the cubic and the orthorhombic phases, the data were refined using two phases. Fig. 6 (d) shows how the fraction of orthorhombic  and cubic phases changes with $T$. As expected, the cubic  fraction diminishes rapidly on cooling and disappears below 200 K. The orthorhombic phase fraction, on the other hand, increases monotonically and almost saturates below 150 K. It is to be noted that the thermal hysteresis in the magnetization and $\mu$SR data  disappears below 150 K. Therefore, 150 K can be assigned as the culminating point of MPT, below which the system attains a stable martensite fraction.  

\begin{table*}
			\begin{tabular}{ p{1.5cm} p{1.5cm} p{1.5cm} p{1.5cm} p{1.5cm}  p{1.5cm}}
					\hline \hline 
 Atom &  Site & X & Y & Z & $B_{iso}$ \\
							\hline 
							      
Ni&        4$h$&     0.0000&     0.249(6)&    0.5000&     1.229(7) \\  
    
Ni&        4$k$&     0.2500&     0.2484(4)&    0.091(5)&     0.544(4)  \\ 
 
Mn&        2$a$&     0.0000&     0.0000&    0.0000&     1.616(6)  \\
  
Mn&        2$f$&     0.2500&     0.5000&    0.574(7)&     1.500(5) \\
    
Mn&        2$b$&     0.0000&     0.5000&    0.0000&     0.990(4)  \\
  
Mn&        2$e$&     0.2500&     0.0000&    0.562(6)&     0.990(4) \\ 
   
Sn&        2$b$&     0.0000&     0.5000&    0.0000&     0.990(4) \\ 
  
Sn&        2$e$&     0.2500&     0.0000&    0.562(6)&     0.990(4) \\ 

\hline 

 \end{tabular}
	\caption{Crystallographic parameter of the sample at 8 K ($Pmma$ ) as obtained from the refinements.}
	\label{ap8}
\end{table*} 

\section{Discussions}
The complex magnetic phases of the studied alloy get reflected in the $\mu$SR data and in association with the NPD result, it clarifies significantly the prevailing doubts on the magnetic states of such Ni-Mn-Z based MSMAs. We  observe that the values of $A_{10}$ obtained  from the $\mu$SR data (both ZF and LF) show a sharp rise on cooling below $T_{MS}$, which continues till the second magnetic transition at $T_{CM}$ is attained. The bulk magnetic measurements [as depicted in figs. 1(a) and (b)] indicate a rapid fall of $M$ below $T_{MS}$, which can be due to the development of a (i) long range ordered AFM state, (ii) long range ordered FM state albeit with highly reduced Mn moment, (iii) a state with short range AFM correlations, (iv) ordered FM clusters in the backdrop of a PM state, or (v) a pure PM state.

\begin{table}
Magnetic moment fixed along [001] direction \\
			$\chi^2$ of fit = 5.45\\
			\begin{tabular}{ p{1cm} p{1cm} p{1cm} p{1cm} p{1cm}  p{1cm}}
					\hline \hline
 Atom &  Site & X & Y & Z & moment ($\mu_B$) \\
							\hline 
							
  &  & 0 & 0 & 0 &  \\ 
 Mn&  2$a$ &  &  &  & 2.761(6) \\
    &  & 0.5 & 0 & 0 &  \\ 
\hline
 &  & 0.25 & 0.5 & 0.574 &  \\ 
 Mn& 2$f$ &  &  &  & 2.302(5) \\ 
    &  & 0.75 & 0.5 & 0.574 &  \\ 

\hline		
		\end{tabular}
		
	\caption{Magnetic structure of the orthorhombic martensite phase ($Pmma$)  as obtained from the powder neutron diffraction measurements at 8 K.}
	\label{mag8}
\end{table} 
\par
This sharp rise in $A_{10}$ below $T_{MS}$ summarily rejects cases (i) and (ii), as long range order should not be accompanied with increasing asymmetry. Therefore, we are left with options (iii), (iv) and (v). If we look at the variation of $A_{10}(T)$, the asymmetry does not fully attain its austenite PM state value just below $T_{MS}$. Therefore, the scenario of a pure PM state can be excluded. The magnetic state just below $T_{MS}$ can be either due to the presence of AFM phase fraction, or associated with the residual FM austenite phase which remained untransformed even below $T_{MS}$. In our NPD data, we find clear signature of this cubic austenite down to 200 K. Therefore the mismatch of $A_{10}$ at 264 K ($\approx T_{CM}$) and 350 K in the ZF $\mu$SR data is likely to be associated with this cubic FM fraction. The variation of  $A_{10} (T)$ in  the LF $\mu$SR data is somewhat similar, although the signature of $T_{MS}$ has shifted slightly to lower temperature. This is  due to the fact that an external magnetic field prefers the ferromagnetically ordered austenite.~\cite{koyama,planes}  Logically, the most probable scenario is case (iv), where the transformed martensite is PM (the major phase), residing along with the untransformed FM austenite (the minor phase).

\par
The thermal hysteresis around the martensitic transition is expected, and it is present in the $\mu$SR data too. Interestingly, both $M(T)$ and $A_{10}(T)$ show another thermal hysteresis between 150 and 225 K. There are several reports on  two-step martensitic transition, where a second inter-martensite transformation occurs.~\cite{fan,mcc} In case of one such  Ni-Mn-In based alloy, the inter-martensite transition was found to occur just below $T_{CM}$, and it was assigned to  a transformation from 10M modulated structure to 14M.~\cite{huang} We  carefully looked at the NPD data in this temperature range, however, no anomaly was detected in the form of peak splitting or appearance of additional reflections. The magnetic transition at around $T_{CM}$ is certainly a first order one, however, it may be an iso-structural one where the lattice symmetry remains unaltered. 

\par
The most important observation in the present work is the signature of $T_B$ in the $\mu$SR data. As evident [see Figs. 3(a) and 4 (a)], the initial asymmetry  shows a rise below $T_B$ indicating the loss of magnetic order in the system. Notably, this rise is present irrespective of the fitting function used (two-exponential or stretched exponential). If we look at the variation of  $\beta$ [see  fig. 5 (b)], it shows a minimum at $T_B$ = 120 K with the  value of $\beta$ close to  0.33. In addition, $\lambda$ shows a weak peak [see  fig. 5 (b)] around $T_B$.  Considering the glassy magnetic state observed in the family of Ni-Mn-Sn alloys below $T_B$, we can  assign $T_B$ to be the spin freezing temperature of the presently studied sample.

\par 
It is now pertinent to discuss the nature and origin of SG ground state. From our NPD data, we observe a single phase orthorhombic martensite at the base temperature. Therefore, the spin-freezing is not related to the presence of minority cubic phase in the system. In Ni-Mn-Z based MSMAs, the sign and strength of magnetic  interaction depend strongly on the Mn-Mn bond distance. Below $T_{MS}$, the intersite Mn-Mn$^{\prime}$ distance decreases paving the path for enhanced AFM correlation.~\cite{montecarlo,vvs-th,pirolkar,parijat} In addition, chemical and lattice disorders play an important role in determining the magnetic ground state of these materials.~\cite{vvs-th}  The  AFM correlations between Mn-Mn$^{\prime}$ (particularly below $T_{MS}$), the Mn-Mn FM  correlations and the presence of disorder eventually lead to spin freezing below $T_B$. The Mn$^{\prime}$atoms are  substituted randomly in the Sn site, which can give rise to  random occurrence of FM and AFM bonds. From  our analysis of the NPD data it is  evident that the AFM correlations is short range in nature, {\it i.e.,} it does not give rise to a long range ordered AFM state. Nevertheless, the ground state does show long range FM order. 
 
\par
In general for a PM to SG transition, the value of $\beta$ in the stretched exponential fitting assumes a constant value below the spin freezing temperature due to the residual fluctuation of the frozen state.~\cite{nav2o5} In contrary to the usual observation, the value of $\beta$ increases below $T_B$ and attains a value of 0.62 at 30 K. The main reason for such anomalous behavior of $\beta$ lies with the  reentrant character of the SG state, where spin freezing takes place on top of a long range FM state. Interestingly, very similar $T$ dependence of $\beta$ was reported in case of Pb(Fe$_{1/2}$Nb$_{1/2}$)O$_3$, where $\beta$ shows a minimum at $T_g$ = 20 K and approaches to unity on further cooling.~\cite{pbfenbo}. The SG state in this compound is  reentrant type which develops in the backdrop of an AFM state. It has been argued that with the critical slowing down of the SG fluctuations at $T_g$, muons starts to sense the non-critical fluctuations of the long range ordered state below $T_g$ leading to the increase in $\beta$. By analogy, the rise in $\beta$ below $T_B$ in the studied Ni-Mn-Sn alloy can be accounted by the presence of long range ordered FM phase with the SG state side by side. This is  corroborated by the NPD data, where FM order is indeed present at 8 K. The weak nature of the peak in $\lambda(T)$ data (ideal spin glass should show a sharp peak at spin freezing temperature) may be due to the presence of this FM ordered state alongside the SG phase.

\par
In conclusion, the present work successfully discerns few ambiguities related to  the magnetic phase diagram of Ni-Mn-Sn alloy system. Based on our study on a particular Ni-Mn-Sn alloy, it is evident that the alloy assumes a PM state just below the martensitic transition. The $\mu$SR result identifies two long range magnetic ordering temperatures $T_{CA}$, and $T_{CM}$ and they are found to be ferromagnetic in nature. The work categorically  justifies the view that the magnetic anomaly at $T_{CM}$ in the martensitic state indeed corresponds to the onset of a long range ordered state. Most importantly, the work identifies that the system transform into a partially disordered magnetic phase below the exchange bias blocking temperature, which can be characterized by the coexistence of ordered FM and frozen spin glass state. The remarkable phenomena of exchange bias observed in Ni-Mn-Z alloys is due to the coupling between the interfacial spins of SG and FM phases.

\section{Acknowledgment}
The work is supported by the India-RAL collaborative project (SR/NM/Z-07/2015). J. Sannigrahi wishes to acknowledge EU’s Horizon 2020 research and innovation programme under the Marie Sk\l{}odowska-Curie Grant Agreement (No. 665593) awarded to the STFC, UK.

\end{document}